\documentclass[runningheads]{llncs}
\pdfoutput=1
\usepackage{tabularx}
\usepackage{graphicx}
\usepackage{adjustbox}
\usepackage{algorithm}
\usepackage{algpseudocode}
\usepackage{amsmath}
\usepackage{multirow}
\usepackage{makecell}
\usepackage[super]{nth}

\usepackage{todonotes}
\usepackage{pgfplots}
\usepackage{pgfplotstable}
\usepgfplotslibrary{groupplots}

\usepackage[utf8]{inputenc}

\usepackage[table]{xcolor}

\pgfplotsset{compat=1.7}
\usepackage{tikz}
\usepackage[square, numbers]{natbib}
\usepackage{comment}
\usepackage{soul}

\usepackage{caption}
\usepackage{subcaption}
\usepackage{booktabs}
\usepackage{makecell}
\usepackage{hyperref}

\makeatletter
\newcommand\avsuminner[2]{%
  {\sbox0{$\m@th#1\sum$}%
   \vphantom{\usebox0}%
   \ooalign{%
     \hidewidth
     \smash{\vrule height\dimexpr\ht0+1pt\relax depth\dimexpr\dp0+1pt\relax}%
     \hidewidth\cr
     $\m@th#1\sum$\cr
   }%
  }%
}
\makeatother

\newcolumntype{P}[1]{>{\centering\arraybackslash}p{#1}}
\newcolumntype{M}[1]{>{\centering\arraybackslash}m{#1}}

\usepackage{enumitem}
\setlist[itemize]{align=parleft,left=0pt..1em}
\usepackage[frozencache,cachedir=minted-cache]{minted}

\definecolor{low}{HTML}{F8696B}    % Kırmızı (Min)
\definecolor{mid}{HTML}{FFEB84}    % Sarı (Sıfır)
\definecolor{high}{HTML}{63BE7B}   % Yeşil (Max)

\pgfplotstableset{
    heatmap/.style={
        col sep=comma,
        string type,
        postproc cell content/.append code={
            \pgfkeysgetvalue{/pgfplots/table/@cell content}\cellcontent
            \ifx\cellcontent\empty
            \else
                \pgfmathparse{IsNumber("\cellcontent")}
                \ifnum\pgfmathresult=1
                    \pgfmathparse{
                        \cellcontent < 0 ? 
                        (100*(\cellcontent/(-32.4))) : % -32.4 tablodaki en düşük değer varsayımı
                        (100*(\cellcontent/32.4))     % 32.4 tablodaki en yüksek değer varsayımı
                    }
                    \edef\temp{\noexpand\lstset{}}
                    \ifdim \cellcontent pt < 0pt
                        \pgfmathsetmacro{\percent}{min(100, max(0, 100*(\cellcontent/-35)))}
                        \edef\cellcolor{\noexpand\cellcolor{low!\percent!mid}}
                    \else
                        \pgfmathsetmacro{\percent}{min(100, max(0, 100*(\cellcontent/35)))}
                        \edef\cellcolor{\noexpand\cellcolor{high!\percent!mid}}
                    \fi
                    \toks0=\expandafter{\cellcolor}
                    \toks1=\expandafter{\cellcontent}
                    \edef\next{\the\toks0 \the\toks1}
                    \pgfkeyslet{/pgfplots/table/@cell content}\next
                \fi
            \fi
        }
    }
}

\begin{document}

\title{Continuous Online Evaluation of Recommendation Strategies in Social Science Academic Search} 
\titlerunning{Continuous Online Evaluation of Recommendations...}

\author{Mehmet Deniz Türkmen \and
Daniel Hienert}

\institute{GESIS - Leibniz Institute for the Social Sciences\\
deniz.tuerkmen@gesis.org\\
daniel.hienert@gesis.org}

%\author{Anonymous Author(s)}

%\institute{Affiliation Omitted for Double-blind Review}

%\authorrunning{Turkmen et al.}
%\authorrunning{M. D.\ Türkmen, and D.\ Hienert}

\maketitle

\begin{abstract}
% What is the problem?
Delivering relevant recommendations in academic search engines is a complex task due to the diversity of subject areas, information types, and user preferences. 
% How do we approach the problem?
In this case study, we address these challenges by integrating and evaluating a range of recommendation systems within GESIS Search -- a domain-specific search engine for the social sciences that provides researchers with access to research data, publications, variables, and measurement instruments. To support continuous, real-time evaluation of multiple recommendation strategies with actual platform users, we utilize the STELLA evaluation framework. We implement and compare a diverse set of algorithms, including traditional lexical similarity, semantic document similarity by using transformer-based embeddings, and session-based recommendations based on click paths from historical user sessions.
% What are the results?
Our results show that users prefer recommendations based on semantic similarity, which outperformed term-similarity and session-based methods. However, the performance of recommenders varies across categories within GESIS Search, suggesting that information-seeking behavior differs by information type.
Overall, our study provides insights into how continuous evaluation can be incorporated to develop recommendations that better align with the preferences in academic search portals.

\begin{comment}
% What is the problem?
It is a challenge to provide the relevant recommendations in academic search engines across different subject areas and information types. Relevant recommendations depend heavily on the subject area, the portal's content, the specific search query, and what the user considers to be a good recommendation and actually clicks on and explores.
% How do we approach the problem?
In this case study, we integrate and evaluate recommendations within a domain-specific search engine for social science information, providing researchers with access to research data, publications, variables, measurement instruments, and more. We incorporate the Stella evaluation framework to enable continuous online evaluation of various recommendation systems using real users in a live environment.
% What are the results?
We have tested different algorithms that provide recommendations based on lexical and semantic document similarity, historical user sessions, and a knowledge graph representing the entire document corpus and its semantic relationships. In a continuous online evaluation, we found that recommendations based on semantic similarity were clearly favored by portal users for this use case. We discuss insights on how continuous evaluation on academic search portals can help identify relevant recommendations for users.
\end{comment}

\end{abstract}

\section{Introduction}
\vspace{-1em}

% Setting the context
Besides large cross-disciplinary search portals for publications~\cite{jacso2005google, lossau2006bielefeld}, there are numerous academic search engines that focus on specific domains, such as bio-medicine \cite{white2020pubmed}, natural sciences~\cite{ginsparg2021lessons}, humanities~\cite{schonfeld2012jstor, valtysson2012europeana}, economics~\cite{ekwurzel1995history,zimmermann2013academic}, and computer science~\cite{white1999acm,ley2002dblp}. These platforms provide access to subject-specific information, such as publications, as well as other types of content, such as research data~\cite{brickley2019google,zenodo_platform}, images, videos \& audio~\cite{valtysson2012europeana}, and research output \& software~\cite{zenodo_platform,hahnel2012exclusive}. Users of these portals are often scientists who rely on the information they find as a basis for their own research~\cite[][p 5]{li2022academic}. Consequently, the topics they search for tend to be highly domain-specific and vary substantially across disciplines~\cite{gasparyan2016specialist}.

% What is the problem?
In addition to ad hoc search, recommendations provide users with suggestions tailored to their information needs~\cite{smeaton2005personalisation}. However, the relevance of recommendations depends on various factors, such as the domain, subject area, type of information, search topics, and the portal's user group. In commercial systems, recommendations are often evaluated directly with the user base, for example, prominently through A/B testing~\cite{nandy2021b}. This allows users to implicitly indicate which recommendations are more relevant through their interactions. In contrast, online evaluation is not yet widely used in academic search engines and is mostly limited to publications~\cite{ong2024cornac}. Recommendations for other information types, such as research data, have been explored sparingly so far, even though recommendations from other researchers seem to play an important role in dataset search~\cite{kramer2021data}.

% How do we approach the problem?
In this case study, we evaluate different types of recommendation strategies in an academic search engine for social science information. To enable continuous online experimentation, we employ the STELLA evaluation framework~\cite{breuer2019stella}, which allows us to compare multiple recommendation systems through interleaving. In this setup, recommendations generated by different systems are combined into a single ranked list, and user interactions are used to infer system preferences. This approach enables continuous deployment and evaluation of new recommendation methods without significantly affecting user experience. 

We evaluate various recommendation approaches based on lexical similarity, semantic similarity, and historical user sessions. Additionally, we perform a category-specific evaluation to better understand how recommender performance varies across different types of scholarly content.
% What are the results?
Our findings show that semantic similarity–based recommenders achieve the strongest overall performance in our use case, outperforming both lexical and session-based methods. However, we also observe that the performance can vary between categories within GESIS Search. In other words, certain recommendation strategies perform better for specific information types, suggesting that users' information needs vary with the nature of the content. 
%We discuss the implications of our findings and how continuous online evaluation can help develop more effective recommendation systems for academic search engines across different domains and use cases with real users.
These findings highlight the importance of continuous online evaluation for developing effective recommendation systems in academic search environments across diverse domains and use cases.

\section{Related Work}\label{sec_rel}

\subsection{Recommendations in Academic Search} 

% recommendation in general
% recommendation in academic search
% different types of recommendations
% evaluation of recommendations

%- Jöreen Beel: Mr. Dlib and co
%- also researchers from siegen
%- review papers!
%    - user preference algorithms for recommenders
%        - subproblem what to choose for a baseline

%While recommender systems are quite common in commercial domains, their use in academic search is comparably limited. Most existing work focuses on recommendations of scholarly publications. One group of studies leverages citation networks to identify related and popular papers. In particular, Küçüktunç et al. \cite{kuccuktuncc2013towards} adapt the PageRank algorithm to citation graphs and incorporate explicit relevance feedback to refine recommendations, while Wang et al. \cite{wang2016academic} extend this idea by additionally using collaboration networks. Other approaches rely on citation patterns more directly: \cite{keshavarz2015parallel} apply locality-sensitive hashing to citation-based signatures in order to efficiently detect papers with similar citation structures.

%While recommender systems are quite common in commercial domains, their use in academic search is comparably limited. On the other hand, modern academic search systems increasingly integrate recommendation components to support scholarly discovery beyond keyword-based retrieval.

Recommender systems have been studied in the field of digital libraries for a long time~\cite{smeaton2005personalisation,gupta2019recommender}. Recommendations in academic search typically focus on related-article suggestions \cite{hanyurwimfura2015effective, beel2016paper}, but have also been applied to venue recommendations \cite{yang2012venue, alhoori2017recommendation, entrup2023comparison, entrup2022b}, personalized research feeds \cite{li2019personalised}, related patent identification \cite{risch2017should, krestel2013recommending}, and scientific video recommendations \cite{medrek2018recommending}. Existing approaches range from content-based techniques to methods that exploit citation networks. In systems with sufficient interaction data, collaborative filtering \cite{hristakeva2017building} and hybrid methods combining textual, citation, and behavioral signals \cite{sterling2021combining} have also been explored.

%citation-based recommendations \cite{kuccuktuncc2013towards, wang2016academic, keshavarz2015parallel},

Content-based approaches typically rely on textual similarity between documents by using representations derived from titles, abstracts, keywords or full texts. Earlier approaches primarily employed TF–IDF-based representations \cite{hurtado2010metadata}, while more recent systems use neural embeddings \cite{kong2018voprec} and transformer-based representations \cite{hassan2019bert} to capture semantic similarity between documents more effectively. Another important line of research focuses on citation-based recommendation methods, which exploit bibliographic networks to identify related work \cite{kuccuktuncc2013towards, wang2016academic, keshavarz2015parallel}. Such approaches are particularly effective in scholarly domains, where citations provide a strong signal of topical and methodological similarities. Similarly, several studies utilize Knowledge Graphs (KGs) to improve recommendations \cite{brack2021citation, wu2024supporting}.

Compared to commercial platforms, personalization in academic search systems remains relatively limited, largely due to privacy concerns and the limited availability of persistent user profiles. Nevertheless, several studies have explored personalized recommendations based on user profiles obtained from publication history, reading behavior, or declared research interests \cite{pera2011personalized, kaur2022feedlens, li2021personalized}. 

In practice, recommendation components in academic search engines are often integrated directly into search result pages or document detail views, where they complement traditional ranked retrieval.
%Academic search recommendations are often embedded directly into search result pages or document detail views, complementing ranked retrieval results. 
Recent work has also explored conversational and generative interfaces that combine retrieval and recommendation to guide literature exploration more interactively \cite{balog2020common, ma2025crs, wang2025llm}.

%Evaluation in this domain gained focus with the development of academic recommendation interfaces. 

Despite numerous works on academic recommender systems, evaluation in this domain is largely limited to offline methods \cite{zangerle2022evaluating, dehghani2019meta, beel2016paper}. On the other hand, several studies argued that test collections fail to reflect real-world user behavior and evolving information needs, motivating a shift toward online evaluation \cite{beel2013comparative, beel2015comparison}. 
%Online evaluation involves user studies and testing in real-world systems. 
Accordingly, Beel et al. \cite{beel2019online} introduced Mr.DLib, a living-lab evaluation framework designed specifically for scholarly recommendation settings. Other studies have further examined how users interact with recommendations and the factors that affect that, such as demographics \cite{beel2013impact}, persistence of displayed results \cite{beel2013persistence}, and material type \cite{charalampous2017classifying}.

\subsection{Continuous Evaluation in Information Retrieval}

Evaluation in information retrieval has traditionally relied on the Cranfield paradigm, where systems are evaluated offline using static document collections, predefined topics, and manually created relevance judgments \cite{cleverdon1967cranfield, voorhees2005trec}. Introduced by shared task campaigns such as TREC \cite{harman1993overview} and CLEF \cite{peters2003cross}, this approach enabled reproducible and standardized comparison between IR systems. 
%However, prior work has noted that static test collections may become obsolete and may not capture evolving corpora, user needs, and retrieval techniques \cite{sanderson2010test, jensen2007repeatable, keller2024evaluation}.
However, prior research has highlighted several limitations of static test collections, particularly their inability to reflect evolving document corpora, changing user needs, and advances in retrieval methods \cite{sanderson2010test, jensen2007repeatable, keller2024evaluation}.

To address these limitations, researchers have explored evaluation approaches that are integrated into live systems. In this context, A/B testing has become a standard method for comparing IR systems based on user interaction signals \cite{kohavi2009controlled}. Interleaving techniques have further improved sensitivity by combining results from different models into a single list and inferring user preferences from click behavior \cite{joachims2002optimizing, radlinski2013optimized}. Together, these approaches have paved the way for continuous system assessment under real usage conditions. 

In addition to industrial applications, the living lab paradigm has also been adopted in academic domains through shared-task campaigns that allow experimental systems to be deployed on operational platforms. \cite{schaer2020living, schaer2021living}. Such environments provide opportunities to evaluate systems directly with real users while maintaining operational search services.

Overall, this body of work reflects a shift from static, one-time assessment toward iterative, deployment-oriented evaluation. 
%Continuous evaluation does not replace offline test collection–based methodologies but complements them by enabling ongoing assessment in dynamic, real-world environments.
Rather than replacing traditional test collection–based evaluation, continuous evaluation complements it by enabling ongoing system assessment in dynamic and real-world environments.

\section{Evaluation and Search Systems}\label{sec_proposed}

In this work, we deployed the STELLA evaluation framework\footnote{https://github.com/stella-project} in the social science search platform GESIS Search\footnote{https://search.gesis.org}. In the following, we outline the principal functionalities of these systems.

\subsection{STELLA Evaluation Framework}

STELLA \cite{breuer2019stella} is an open-source evaluation framework that enables portal operators and researchers to conduct experiments with real users in real-world environments. It provides an Evaluation-as-a-Service platform for living lab experiments involving ranking and recommender systems. 
%Using STELLA, researchers can evaluate their experimental systems based on user feedback, unlike Cranfield-style approaches that rely on test collections in offline evaluations. 
Unlike traditional Cranfield-style evaluation approaches that rely on offline test collections, STELLA facilitates the assessment of experimental systems through user interactions and feedback.
In addition to conventional A/B testing, the framework supports interleaving-based system comparison, where the outputs of two ranking or recommendation models are combined into a single result list for evaluation. STELLA's infrastructure consists of four main components:

\begin{itemize}
    \item[] \textbf{Experimental systems}: A set of search or recommender systems to be evaluated within the framework.
    \item[] \textbf{Site}: Platform where experimental systems interact with real users.
    \item[] \textbf{STELLA Server}: The administrative component of the infrastructure, responsible for system management, dashboard access, storage of user interaction logs, and visualization of evaluation results.
    \item[] \textbf{STELLA App}: A deployable application installed on participating platforms to conduct retrieval and recommender system experiments using real user interactions. The app is mainly responsible for 
    receiving rankings or recommendations from experimental systems, preparing the result list in line with the evaluation technique (i.e., A/B testing or interleaving), and conveying the result to the platform. It also collects interaction data, which is subsequently transmitted to the STELLA Server.
\end{itemize}

\subsection{Social Science Academic Search} \label{subsec:gesis_search}

GESIS Search \cite{hienert2019digital} is an academic search platform that enables researchers to access and explore a broad range of social science resources within a single system. The platform supports cross-category search over research datasets, publications, survey variables, questionnaire items, survey instruments and tools, as well as webpages and publications from the GESIS library. In addition to keyword-based retrieval, the platform provides semantic links between information objects, allowing users to explore relationships between resources. For example, users can identify publications that cite specific research datasets or list variables associated with a particular study. To help researchers perform targeted searches, GESIS Search offers advanced filtering based on study collections and metadata attributes. The platform also includes functions for downloading datasets and related materials, which facilitate data reuse. Overall, the platform contains six categories of information. In this case study, recommendations are implemented for all categories except GESIS webpages.

%GESIS Search \cite{hienert2019digital} enables researchers to find social science information on research data and related information in one place. Users can search across categories for research datasets, publications, survey variables, questionnaire questions, survey instruments and tools, as well as GESIS webpages and publications from the GESIS library. Information items are linked to each other, enabling users to see, for example, which publications contain data citations to research data. 

%The platform offers advanced search filters that enable refined searches across different study collections and study attributes. The ability to download data and materials allows researchers to reuse data to answer their own research questions. There are six categories of information available on the platform. In this case study, we are implementing recommendations for all categories except the GESIS webpages.

\begin{itemize}
    \item[] \textbf{Publications}: $\sim$260,000 publications from open-access sources and literature linked to research datasets.
    \item[] \textbf{Research Data}: $\sim$7,800 quantitative social science datasets covering Germany and Europe from 1945 to 2026, provided through the GESIS Data Archive.
    \item[] \textbf{Variables}: $\sim$1.4 million survey variables extracted from questionnaires. Researchers can access high-quality variables with question texts, answer categories, and frequency tables.
    \item[] \textbf{Instrument \& Tools}: $\sim$620 resources related to survey design and data collection, such as pretested questionnaires, response scales, syntax files for analysis in statistical software, and guidelines for survey and behavioral data collection.
    \item[] \textbf{GESIS Library}: $\sim$122,000 publications from the GESIS library, with a focus on empirical social science and applied computer science.
    \item[] \textbf{GESIS Webpages}: $\sim$2,700 webpages from the GESIS website, including training materials, consulting services, and institutional information for social science researchers.
\end{itemize}

\section{Experimental Recommender Systems}

%In this section, we describe the different experimental recommender systems we implemented and used in the online evaluation.
In this section, we describe the recommender systems implemented and evaluated within the online living lab environment.

\subsection{Term Similarity}

All documents in the Gesis Search index follow a shared metadata schema consisting of common fields such as title and abstract, complemented by type-specific metadata standards (e.g., DDI metadata \footnote{https://ddialliance.org/} for research datasets). The term similarity recommender identifies related documents based on lexical overlap between these fields and those of a given source document.
%All documents in the search index use a common metadata schema with fields such as title and abstract supplemented by individual schemas for different information types, such as DDI\footnote{https://ddialliance.org/} for research data. The term similarity recommender proposes documents that contain similar terms to a given source document.
To implement this approach, we rely on the “more like this” (MLT) query in Elasticsearch \footnote{https://www.elastic.co/docs/reference/query-languages/query-dsl/query-dsl-mlt-query}. The MLT query extracts text from the source document based on the field mapping. It then selects the top-k terms with the highest TF-IDF scores. These terms are used in a disjunctive query to retrieve and recommend the most similar documents from the index.
%For implementation of this approach, we use Elasticsearch's "more\_like\_this" (MLT) query\footnote{https://www.elastic.co/docs/reference/query-languages/query-dsl/query-dsl-mlt-query} that can be customized with a number of parameters. The basic idea is that documents can be well represented by their TF-IDF values.
%Essentially, in the MLT query, we reference the input document from the index. Then it extracts the text from the input document and analyzes it with the fields analyzer given in the field mapping. This builds a set of top-k TD-IDF scores representing the source document. These terms are then used in a disjunctive query to find similar target documents from the index. Target documents are then used as recommendations.

Listing \ref{listing:more-like-this} shows the instantiated MLT query. The ``fields'' parameter serves three purposes: (1) It determines which fields are used for text extraction from the source document and for querying candidate documents. (2) It allows weighting of fields; for example, terms appearing in titles can receive higher weights than terms in other fields. (3) It links fields to specific analyzers defined in the index mapping. For example, the field \texttt{title.partial} applies lowercasing and n-gram tokenization with token lengths between four and eight characters. This allows the query to find partial matches and retrieval of related terms such as ``migration'', ``migrant'', or ``emigration''. Overall, the field configuration provides detailed control over term extraction, field weights, and linguistic analysis.
%This creates subterms from given source terms to find target documents that contain them, e.g., the subterm 'migra' to find 'emigration' and 'migrant'. Altogether, the fields definition lets us define from where terms are extracted, how they are weighted, and how they are analyzed. 

\begin{listing}[!ht]
\begin{minted}[fontsize=\scriptsize]{python}
"query": {
  "more_like_this": {
    "fields": [
      "_all","title^10","topic^7","abstract^3","title_en^10","topic_en^7","abstract_en^3",
      "title.partial^0.4","topic.partial^0.3","abstract.partial^0.2","content.partial^0.4",
      "full_text^0.1"
    ],
    "like": [
      {
        "_index": index_name,
        "_id": item_id
      }
    ]
  }
}
\end{minted}
\caption{The more-like-this query used for querying recommendations with similar terms.}
\label{listing:more-like-this}
\end{listing}

\subsection{Semantic Similarity}

To explore beyond term-matching, we implemented recommenders based on semantic similarity using text embeddings. Unlike the term-similarity approach relying on lexical overlap, this technique captures the meaning of the text by representing documents as dense vectors.
%While term-based approaches rely on lexical overlap, semantic models focus on semantic similarity, capturing the meaning of texts.
%Text embeddings are known to capture the meaning of text. Accordingly, we employ embedding similarity as an alternative to term similarity. 
To implement semantic recommendations, we first extract and concatenate textual information from selected metadata fields, such as abstract and title (Table~\ref{tab:fields-for-embeddings}), to form a textual representation for each scholarly item. Then, these are encoded into dense vectors using various embedding models. 
%For each item, we extracted and concatenated texts from selected fields, such as title and abstract (see Table~\ref{tab:fields-for-embeddings}) to form a textual feature for each document. Then, we fed textual features to Sentence Transformer models to compute dense semantic representations of items. 

In this study, we experimented with three embedding models: (1) \textit{para\-phrase-multilingual-mpnet-base-v2} \cite{reimers-2019-sentence-bert}, (2) \textit{nomic-embed-text-v2-moe} ~\cite{nussbaum2025trainingsparsemixtureexperts}, and (3)
\textit{all-MiniLM-L6-v2}\footnote{https://huggingface.co/sentence-transformers/all-MiniLM-L6-v2}. These models were selected because they are multilingual and support the document corpus with a mixture of German and English documents and field contents. The first two models produce embedding vectors of length 768, whereas the third produces vectors of length 384.

The resulting vectors are indexed in Elasticsearch as dense vectors. During retrieval, we use Elasticsearch’s K-nearest neighbor (KNN) query to identify semantically similar items based on cosine similarity between embedding vectors. Candidate documents with cosine similarity scores above 0.8 are retained as recommendations.
%Since the KNN implementation performs approximate nearest neighbor (ANN) search, it enables efficient retrieval in large vector spaces.
%Elasticsearch’s KNN query is used for approximate nearest neighbor (ANN) search on vector fields, which efficiently finds similar items. Finally, we filter documents with a cosine similarity above 0.8.

\begin{table}[]
\small
\centering
\begin{tabular}{ll}
\hline
\textbf{Information Category} & \textbf{Fields} \\ \hline
Research Data & title, abstract, content\_description \\
Variables & title, question\_text \\
Instruments \& Tools & title, abstract, topic, theory \\
Publications & title, abstract, topic \\
GESIS Library & title, abstract, topic \\ \hline
\end{tabular}
\caption{Fields per information category used to build textual features for embeddings.}
\label{tab:fields-for-embeddings}
\end{table}

Listing \ref{listing:knn} shows an example KNN query. In the query, the ``field'' parameter is the field name where the embedding vector is stored in the ES index, while ``query\_vector'' is the embedding vector itself of the target item. ``k'' represents the number of similar items to be found. ``num\_candidates'' is the number of candidates to be explored before selecting the top-k items. 
%While larger values improve search accuracy, they also increase the cost. 
Higher values improve retrieval quality at the cost of increased computational overhead.
%For this parameter, we chose 100 to balance search accuracy and cost. Finally, we limit the search space to items of the same type as the source item.
In our implementation, we set this value to 100 to balance efficiency and effectiveness. In addition, retrieval is restricted to items belonging to the same information category as the source item.

\begin{listing}[!ht]
\begin{minted}[fontsize=\scriptsize]{python}
"knn": {
    "field": embedding_field,
    "query_vector": source_item_embedding,
    "k": k,
    "num_candidates": 100,
    "filter": {"term": {"type": source_item_type}}
}
\end{minted}
\caption{The knn query used for retrieving based on embedding similarity.}
\label{listing:knn}
\end{listing}

%\subsection{Session-based RNN Recommender}
\subsection{Session-based Click-Path Recommender}

We implemented a session-based recommender that predicts users’ next interaction based on anonymous click-path sequences in historical data.
%The session-based recommender models the likelihood of a user's next item view based on all past users with similar item views in their sessions. Therefore, we developed a recommender that predicts the next item a user is likely to click, given the sequence of items clicked within the same session. 
Unlike collaborative filtering approaches that require persistent user profiles, this method relies on interactions within the current session and therefore supports anonymous recommendations.
The model was trained on user interaction logs that span eight years, comprising 16M user interactions. %The recommender does not require personal user profiles; instead, it relies on current session interactions to make anonymous recommendations. Moreover, it can work with sparse input data, unlike collaborative filtering methods. We represent each search session as a sequence of user clicks (click path) related to a specific item:
Each session is represented as an ordered sequence of item interactions:

\[s = (i_1, i_2, \ldots, i_T) \]
where $i_t$ denotes the item interacted at position $t$ in the session and $T$ is the session length. 

The interaction sequence includes several types of user actions:
%The user actions that are considered for the click-path list are
(1) opening the detailed view of an item, (2) export actions such as exporting a citation, (3) dataset and questionnaire downloads, and (4) bookmarking items to the watchlist. Each of these actions indicates a specific user interest in an item. Every time the user changes their query, the click-path sequence is reset.

Given the subsequence $(i_1,i_2,...,i_{t})$, the task is to predict the next click $i_{t+1}$. 
Each item is represented with an embedding vector $e(i)$ of 768 length, which is generated using \textit{nomic-embed-text-v2-moe} model~\cite{nussbaum2025trainingsparsemixtureexperts}. To model sequential behavior, we employ a Recurrent Neural Network (RNN) that processes item embeddings in chronological order. Specifically, we use a single Gated Recurrent Unit (GRU) layer with 1000 hidden units to encode the current session's content:

\[h_t = GRU(h_{t-1}), e(i_t).\]
The final hidden state ($h_t$) of the GRU is then passed through a fully connected layer with 512 units to predict the embedding of the next likely click.

\[\hat{e}_{t+1} = FC(h_t).\]
The model is trained by minimizing cosine similarity loss between the predicted embedding and the embedding of the true next-clicked item:

\[L = 1 - cos(\hat{e}_{t+1}, e_{t+1})).\]
To expand the size of the training data and to improve robustness for short sessions, we apply prefix augmentation. In particular, for each original session $s=(i_1,i_2,...,i_{T})$, all prefixes

\[ (i_1),(i_1,i_2),...,(i_1,...,i_{T-1}) \]
are treated as independent training samples with their corresponding next-item targets $i_2,i_3,...,i_T$. 

At inference time, recommendations are generated through a KNN search in the embedding space using cosine similarity between the predicted embedding and candidate item embeddings. The top-k items with the highest similarity are returned as recommendations.

\section{Experiments} \label{sec_exp_real}

In this section, we describe the evaluation use case, experimental methods used, and the results.

\subsection{Use Case: Item Recommendation in GESIS Search}

\subsubsection{Use Case Overview:}
\begin{figure}
\centering
\frame{\includegraphics[width=1\textwidth]{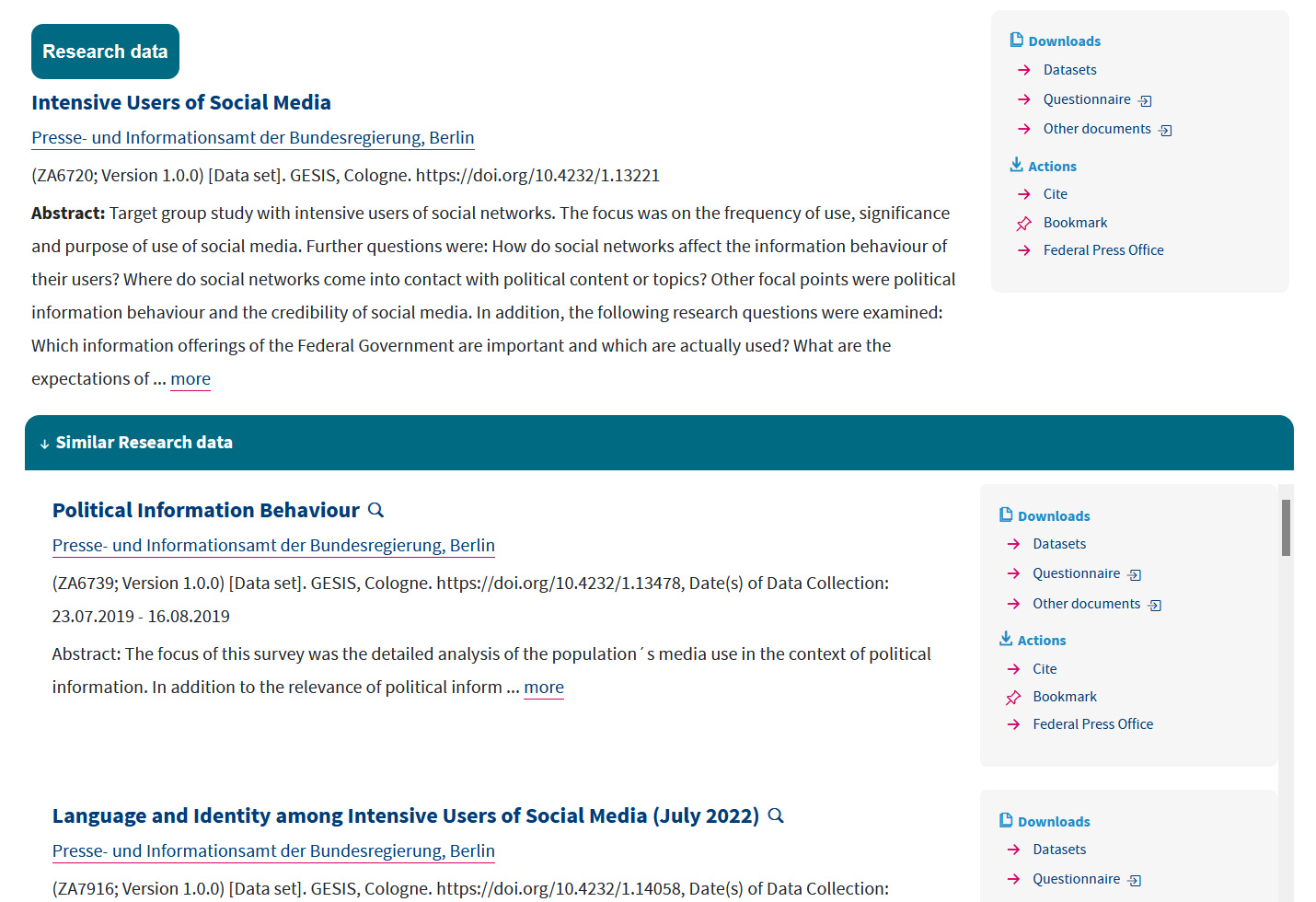}}
\caption{Recommendations in the section \textit{Similar Research Data} for a given research dataset about social media usage.} \label{fig:research_recommender}
\end{figure}

This study focuses on the development and evaluation of recommendation systems for five categories in GESIS Search, namely Research Data, Variables, Instruments \& Tools, Publications, and GESIS Library.
To assess the effectiveness of various approaches in a real-world environment, we utilize the STELLA framework, which enables live evaluation with active GESIS Search users. The recommendation section is integrated into the detailed view of each record. It displays up to 10 items in the same category as the currently viewed item; for instance, a user viewing a research dataset will see a list of similar datasets, as shown in Figure \ref{fig:research_recommender}.

\subsubsection{Evaluation:}
To determine the best approach for our use case, we implemented five recommender systems representing three distinct algorithmic classes. These systems were deployed as Docker containers and registered within the STELLA App. The evaluation is based on the interleaving method. When a user views an item, two candidate systems concurrently generate recommendation lists. The STELLA App merges these lists into a single interleaved list, which is displayed to the user. User clicks on recommended items are recorded and attributed to the corresponding recommendation system. After a defined time period, the system with more clicks is considered the winner of the comparison.

\subsubsection{User Interface and Interaction:}
Recommendations are displayed in the recommendations section as a standard hit list, allowing users to assess relevance based on titles, authors, and abstracts. Additional actions, including data access, full-text links, and citation export functions, are displayed alongside the recommendations. Although the exact layout and the specific metadata displayed vary slightly by item type (e.g., variables vs. publications), the core functionality is consistent. Clicking on a title opens the full view, where all metadata and available functions can be explored.  Overall, this interface provides a natural context for users to interact with recommendations during exploratory search \cite{hienert2019digital}.

\subsection{Experimental Method}\label{sec_exp_setup}

We evaluated five recommender systems using interleaving on the GESIS Search platform. Interleaving enables sensitive pairwise comparisons by combining the output of two systems into a single result list. %Since interleaving supports only pairwise comparisons, 
Accordingly, we compared each pair of systems, resulting in a complete set of pairwise evaluations across all systems.

For all comparisons, we used team-draft interleaving (TDI) \cite{radlinski2008does}, which creates a single result list by selecting items from the two systems in a randomized draft order, minimizing bias. User interactions with the interleaved rankings were interpreted as implicit preferences for one of the two systems.

Because interleaving yields paired binary outcomes indicating which system wins for each impression, we assessed statistical significance using a two-sided binomial sign test under the null hypothesis of equal user preference. In order to compare multiple system pairs, we used the Holm–Bonferroni correction with a significance level of $\alpha=0.01$.

\subsection{Results}

Table \ref{tab:pairwise_system_comp} reports the results from the interleaving experiments for each system pair. The table contains the number of wins for each system of the pair and the p-value indicating the significance of the comparison. Significantly better performing systems are highlighted in bold. The results show that the baseline recommender based on term similarity received fewer clicks than other approaches, demonstrating the superiority of embedding-based recommendation methods. Although the click-path recommender (CPR) surpassed the term-similarity baseline, it remained less effective than the embedding-based similarity models. 

The comparisons between embedding-based similarity models demonstrate that the nomic model produced the most favorable recommendations for GESIS Search users, while recommenders using the MiniLM and mpnet-base models performed very similarly, with no significant difference. According to the results, the overall ranking of recommender effectiveness can be summarized as follows: nomic $\succ$ MiniLM $\equiv$ mpnet-base $\succ$ CPR $\succ$ term-similarity.

We further analyzed the recommendation performance of systems across scholarly categories of GESIS Search. To this end, we identified the clicked item categories and calculated the win-lose counts for each category separately (see \textbf{Table \ref{tab:category-specific}}). The table clearly shows that recommendations for Publications received far more clicks than the other scholarly types. In contrast, very few clicks are directed to Instruments \& Tools category. The term-similarity recommender performed particularly poorly for Publication category compared to its overall performance, as indicated by the shades of red in the table (i.e., rows 1, 2, 3, and 7). In contrast, the same approach achieved noticeably stronger results for Research Data and Variable categories, where its performance improved by approximately 15-20\% relative to the overall results. In these categories, the term-similarity recommender even outperformed the click-path recommender. A similar improvement in these two categories was also observed for CPR. Among all categories, the GESIS Library results aligned most closely with the overall evaluation results from Table \ref{tab:pairwise_system_comp}.

\begin{table*}[!htbp]
%\small
\centering
    \begin{tabular}{lM{2.5cm}M{2.5cm}cc}
    \toprule
    \makecell{\textbf{Comparison Pair}\\$A \succ B$} & \textbf{A wins} & \textbf{B wins} & \textbf{\#queries} & \textbf{p-value}                \\ 
    \toprule
    %Phase 1 & & & \\
    term-sim $\succ$ \textbf{nomic}      & 271 (25.4\%)          & \textbf{797 (74.6\%)} & 1068  & \textbf{4.5x10\textsuperscript{-15}}  \\
    term-sim $\succ$ \textbf{minilm}     & 340 (30.8\%)          & \textbf{765 (69.2\%)} & 1105  & \textbf{3.9x10\textsuperscript{-15}}  \\
    term-sim $\succ$ \textbf{mpnet-base} & 312 (29.8\%)          & \textbf{735 (70.2\%)} & 1047  & \textbf{5x10\textsuperscript{-15}}    \\
    \midrule
    %\hline
    \textbf{nomic} $\succ$ minilm        & \textbf{769 (56.6\%)} & 590 (43.4\%)          & 1359  & \textbf{1.3x10\textsuperscript{-6}}   \\
    \textbf{nomic} $\succ$ mpnet-base    & \textbf{767 (55.5\%)} & 615 (44.5\%)          & 1382  & \textbf{4.8x10\textsuperscript{-5}}   \\
    mpnet-base $\succ$ minilm            & 606 (49.7\%)          & 613 (50.3\%)          & 1219  & 0.84                                  \\
    \midrule
    \textbf{CPR} $\succ$ term-sim        & \textbf{602 (54.9\%)} & 495 (45.1\%)          & 1097  & \textbf{0.0014}                       \\
    CPR $\succ$ \textbf{nomic}           & 339 (28.7\%)          & \textbf{844 (71.3\%)} & 1183  & \textbf{5.1x10\textsuperscript{-15}}  \\
    CPR $\succ$ \textbf{minilm}          & 349 (29.3\%)          & \textbf{844 (70.7\%)} & 1193  & \textbf{4.1x10\textsuperscript{-15}}  \\
    CPR $\succ$ \textbf{mpnet-base}      & 344 (30.9\%)          & \textbf{768 (69.1\%)} & 1112  & \textbf{3.8x10\textsuperscript{-15}}  \\
  
    \bottomrule

    \end{tabular}
    \caption{Pairwise comparison of recommender systems. Statistically significant p-values are highlighted in bold.}
    \label{tab:pairwise_system_comp}
\end{table*}

\begin{table}[h]
    \centering
    \tiny
    \begin{tabular}{lcc|cc|cc|cc|cc}
        \toprule
        & \multicolumn{2}{c}{Publications} & \multicolumn{2}{c}{Research data} & \multicolumn{2}{c}{Variables} & \multicolumn{2}{c}{Instruments} & \multicolumn{2}{c}{GESIS Lib} \\
        \midrule
\multirow{2}{*}{term-sim X nomic} & \cellcolor{low!55!mid}85 & \cellcolor{high!55!mid}501 & \cellcolor{high!82!mid}138 & \cellcolor{low!82!mid}192 & \cellcolor{high!58!mid}33 & \cellcolor{low!58!mid}56 & \cellcolor{low!2!mid}2 & \cellcolor{high!2!mid}6 & \cellcolor{low!9!mid}13 & \cellcolor{high!9!mid}42 \\
& \cellcolor{low!55!mid}(14.5\%) & \cellcolor{high!55!mid}(85.5\%) & \cellcolor{high!82!mid}(41.8\%) & \cellcolor{low!82!mid}(58.2\%) & \cellcolor{high!58!mid}(37.1\%) & \cellcolor{low!58!mid}(62.9\%) & \cellcolor{low!2!mid}(25.0\%) & \cellcolor{high!2!mid}(75.0\%) & \cellcolor{low!9!mid}(23.6\%) & \cellcolor{high!9!mid}(76.4\%) \\
& & & & & & & & & &  \\
\multirow{2}{*}{term-sim X minilm} & \cellcolor{low!57!mid}118 & \cellcolor{high!57!mid}488 & \cellcolor{high!86!mid}156 & \cellcolor{low!86!mid}169 & \cellcolor{high!66!mid}43 & \cellcolor{low!66!mid}55 & \cellcolor{high!74!mid}5 & \cellcolor{low!74!mid}6 & \cellcolor{low!16!mid}18 & \cellcolor{high!16!mid}47 \\
& \cellcolor{low!57!mid}(19.5\%) & \cellcolor{high!57!mid}(80.5\%) & \cellcolor{high!86!mid}(48.0\%) & \cellcolor{low!86!mid}(52.0\%) & \cellcolor{high!66!mid}(43.9\%) & \cellcolor{low!66!mid}(56.1\%) & \cellcolor{high!74!mid}(45.5\%) & \cellcolor{low!74!mid}(54.5\%) & \cellcolor{low!16!mid}(27.7\%) & \cellcolor{high!16!mid}(72.3\%) \\
& & & & & & & & & &  \\
\multirow{2}{*}{term-sim X mpnet-base} & \cellcolor{low!60!mid}94 & \cellcolor{high!60!mid}430 & \cellcolor{high!72!mid}155 & \cellcolor{low!72!mid}196 & \cellcolor{high!46!mid}36 & \cellcolor{low!46!mid}56 & \cellcolor{high!66!mid}3 & \cellcolor{low!66!mid}4 & \cellcolor{high!16!mid}24 & \cellcolor{low!16!mid}49 \\
& \cellcolor{low!60!mid}(17.9\%) & \cellcolor{high!60!mid}(82.1\%) & \cellcolor{high!72!mid}(44.2\%) & \cellcolor{low!72!mid}(55.8\%) & \cellcolor{high!46!mid}(39.1\%) & \cellcolor{low!46!mid}(60.9\%) & \cellcolor{high!66!mid}(42.9\%) & \cellcolor{low!66!mid}(57.1\%) & \cellcolor{high!16!mid}(32.9\%) & \cellcolor{low!16!mid}(67.1\%) \\
& & & & & & & & & &  \\
\multirow{2}{*}{nomic X minilm} & \cellcolor{high!6!mid}484 & \cellcolor{low!6!mid}354 & \cellcolor{low!16!mid}162 & \cellcolor{high!16!mid}141 & \cellcolor{high!41!mid}46 & \cellcolor{low!41!mid}25 & \cellcolor{high!30!mid}5 & \cellcolor{low!30!mid}3 & \cellcolor{low!24!mid}72 & \cellcolor{high!24!mid}67 \\
& \cellcolor{high!6!mid}(57.8\%) & \cellcolor{low!6!mid}(42.2\%) & \cellcolor{low!16!mid}(53.5\%) & \cellcolor{high!16!mid}(46.5\%) & \cellcolor{high!41!mid}(64.8\%) & \cellcolor{low!41!mid}(35.2\%) & \cellcolor{high!30!mid}(62.5\%) & \cellcolor{low!30!mid}(37.5\%) & \cellcolor{low!24!mid}(51.8\%) & \cellcolor{high!24!mid}(48.2\%) \\
& & & & & & & & & &  \\
\multirow{2}{*}{nomic X mpnet-base} & \cellcolor{high!9!mid}492 & \cellcolor{low!9!mid}366 & \cellcolor{low!26!mid}162 & \cellcolor{high!26!mid}161 & \cellcolor{low!31!mid}37 & \cellcolor{high!31!mid}38 & \cellcolor{high!98!mid}6 & \cellcolor{low!98!mid}2 & \cellcolor{high!19!mid}70 & \cellcolor{low!19!mid}48 \\
& \cellcolor{high!9!mid}(57.3\%) & \cellcolor{low!9!mid}(42.7\%) & \cellcolor{low!26!mid}(50.2\%) & \cellcolor{high!26!mid}(49.8\%) & \cellcolor{low!31!mid}(49.3\%) & \cellcolor{high!31!mid}(50.7\%) & \cellcolor{high!98!mid}(75.0\%) & \cellcolor{low!98!mid}(25.0\%) & \cellcolor{high!19!mid}(59.3\%) & \cellcolor{low!19!mid}(40.7\%) \\
& & & & & & & & & &  \\
\multirow{2}{*}{mpnet-base X minilm} & \cellcolor{low!3!mid}396 & \cellcolor{high!3!mid}410 & \cellcolor{high!27!mid}129 & \cellcolor{low!27!mid}105 & \cellcolor{low!8!mid}24 & \cellcolor{high!8!mid}26 & \cellcolor{high!52!mid}3 & \cellcolor{low!52!mid}2 & \cellcolor{low!31!mid}54 & \cellcolor{high!31!mid}70 \\
& \cellcolor{low!3!mid}(49.1\%) & \cellcolor{high!3!mid}(50.9\%) & \cellcolor{high!27!mid}(55.1\%) & \cellcolor{low!27!mid}(44.9\%) & \cellcolor{low!8!mid}(48.0\%) & \cellcolor{high!8!mid}(52.0\%) & \cellcolor{high!52!mid}(60.0\%) & \cellcolor{low!52!mid}(40.0\%) & \cellcolor{low!31!mid}(43.5\%) & \cellcolor{high!31!mid}(56.5\%) \\
& & & & & & & & & &  \\
\multirow{2}{*}{CPR X term-sim} & \cellcolor{high!66!mid}319 & \cellcolor{low!66!mid}150 & \cellcolor{low!68!mid}187 & \cellcolor{high!68!mid}267 & \cellcolor{low!30!mid}42 & \cellcolor{high!30!mid}44 & \cellcolor{high!100!mid}10 & \cellcolor{low!100!mid}3 & \cellcolor{high!19!mid}44 & \cellcolor{low!19!mid}31 \\
& \cellcolor{high!66!mid}(68.0\%) & \cellcolor{low!66!mid}(32.0\%) & \cellcolor{low!68!mid}(41.2\%) & \cellcolor{high!68!mid}(58.8\%) & \cellcolor{low!30!mid}(48.8\%) & \cellcolor{high!30!mid}(51.2\%) & \cellcolor{high!100!mid}(76.9\%) & \cellcolor{low!100!mid}(23.1\%) & \cellcolor{high!19!mid}(58.7\%) & \cellcolor{low!19!mid}(41.3\%) \\
& & & & & & & & & &  \\
\multirow{2}{*}{CPR X nomic} & \cellcolor{low!16!mid}181 & \cellcolor{high!16!mid}529 & \cellcolor{high!26!mid}89 & \cellcolor{low!26!mid}173 & \cellcolor{high!55!mid}27 & \cellcolor{low!55!mid}41 & \cellcolor{high!57!mid}6 & \cellcolor{low!57!mid}9 & \cellcolor{low!3!mid}36 & \cellcolor{high!3!mid}92 \\
& \cellcolor{low!16!mid}(25.5\%) & \cellcolor{high!16!mid}(74.5\%) & \cellcolor{high!26!mid}(34.0\%) & \cellcolor{low!26!mid}(66.0\%) & \cellcolor{high!55!mid}(39.7\%) & \cellcolor{low!55!mid}(60.3\%) & \cellcolor{high!57!mid}(40.0\%) & \cellcolor{low!57!mid}(60.0\%) & \cellcolor{low!3!mid}(28.1\%) & \cellcolor{high!3!mid}(71.9\%) \\
& & & & & & & & & &  \\
\multirow{2}{*}{CPR X minilm} & \cellcolor{low!14!mid}187 & \cellcolor{high!14!mid}522 & \cellcolor{high!50!mid}117 & \cellcolor{low!50!mid}180 & \cellcolor{high!3!mid}18 & \cellcolor{low!3!mid}42 & \cellcolor{high!30!mid}6 & \cellcolor{low!30!mid}11 & \cellcolor{low!51!mid}21 & \cellcolor{high!51!mid}89 \\
& \cellcolor{low!14!mid}(26.4\%) & \cellcolor{high!14!mid}(73.6\%) & \cellcolor{high!50!mid}(39.4\%) & \cellcolor{low!50!mid}(60.6\%) & \cellcolor{high!3!mid}(30.0\%) & \cellcolor{low!3!mid}(70.0\%) & \cellcolor{high!30!mid}(35.3\%) & \cellcolor{low!30!mid}(64.7\%) & \cellcolor{low!51!mid}(19.1\%) & \cellcolor{high!51!mid}(80.9\%) \\
& & & & & & & & & &  \\
\multirow{2}{*}{CPR X mpnet-base} & \cellcolor{low!15!mid}179 & \cellcolor{high!15!mid}462 & \cellcolor{high!38!mid}110 & \cellcolor{low!38!mid}176 & \cellcolor{high!12!mid}20 & \cellcolor{low!12!mid}40 & \cellcolor{high!100!mid}7 & \cellcolor{low!100!mid}4 & \cellcolor{low!32!mid}28 & \cellcolor{high!32!mid}86 \\
& \cellcolor{low!15!mid}(27.9\%) & \cellcolor{high!15!mid}(72.1\%) & \cellcolor{high!38!mid}(38.5\%) & \cellcolor{low!38!mid}(61.5\%) & \cellcolor{high!12!mid}(33.3\%) & \cellcolor{low!12!mid}(66.7\%) & \cellcolor{high!100!mid}(63.6\%) & \cellcolor{low!100!mid}(36.4\%) & \cellcolor{low!32!mid}(24.6\%) & \cellcolor{high!32!mid}(75.4\%) \\

        \bottomrule
    \end{tabular}
    \caption{Category-specific pairwise comparison of recommender systems. For each pair and category, two cells contain the number of wins and the win rate for systems in the corresponding pair. Cell colors indicate percentage change from the overall results (i.e., Table \ref{tab:pairwise_system_comp}): decreases (red), increases (green), and near-zero changes (yellow).}
    \label{tab:category-specific}
\end{table}

\section{Discussion and Limitations} \label{sec_disc} \label{sec_limit}

% Intro to discussion: details on recommender click rate and fewer clicks per time unit for lower-quality recommenders
In our evaluation framework, we compared five recommender systems through ten pairwise comparisons, collecting at least 1,000 clicks per comparison and 11,765 clicks in total. Given the current daily traffic of GESIS Search with 30k document views per week, we observed an average of around 500 clicks per week in the recently introduced recommendation section. Consequently, evaluating all five systems required approximately 23 weeks. We also observed that comparisons involving stronger systems accumulated clicks more quickly, whereas experiments with weaker recommenders required substantially more time to reach the same threshold. This suggests that users are less likely to engage with low-quality recommendations and may ignore them altogether. 

%Transitivity between recommender systems is a detail of the discussion intro/overview
%Our evaluation results satisfy the transitivity property: for any three systems $A$, $B$, and $C$ in our system set, if $A > B$ and $B > C$, then $A > C$. This indicates our online evaluation setup produces consistent and reliable rankings. An important implication is that a set of n systems can be evaluated and ranked with fewer comparisons (i.e., \( O(n \log n) \)), rather than exhaustively comparing all pairs (i.e., \( \binom{n}{2} \)), which greatly reduces evaluation cost.
Our evaluation results satisfy Strong Stochastic Transitivity (SST)\cite{kozielecki1982psychological}: for any three systems $A$, $B$, and $C$ in our set, if the win probabilities $P(A, B) \geq 0.5$ and $P(B, C) \geq 0.5$, then $P(A, C) \geq \max[P(A, B), P(B, C)]$. This property indicates that our online evaluation setup produces consistent rankings, and the magnitude of preference increases as the performance gap between systems grows. From a practical point of view, this property confirms that a set of $n$ systems can be reliably ranked with fewer comparisons (i.e., \( O(n \log n) \)), rather than exhaustively comparing all pairs (i.e., \( \binom{n}{2} \)). This substantially reduces the total evaluation cost and user traffic required to determine system ranking.

% Results for term-based similarity: worst, but has been used prominently in (academic search engines)
% discuss on first position here, because it is the baseline and the simplest approach
The term-based similarity recommender is the simplest approach among the five models and serves as the baseline in our experiments. Because it relies on exact-term matching (with n-gram analysis), it cannot capture semantic similarity between documents with different wording but the same concepts. As a result, it is better suited to known-item search scenarios (e.g., dataset search \cite{carevic2020characteristics}) than to exploratory search. In line with earlier work \cite{frasincar2012semantic, capelle2012semantics}, our results show that term-based methods are outperformed by semantic similarity approaches in exploratory recommendation settings.
%It relies on a classical approach that has been widely trusted and used across many domains and platforms for decades. Our results show that, while it still meets users' needs for recommendations, it receives far fewer clicks than recommendations based on semantic similarity. 
Nevertheless, this approach has clear strengths in terms of efficiency. It operates directly on existing textual metadata without requiring vector computation, storage, or approximate nearest neighbor search, all of which can be computationally demanding. Especially in large systems with frequently updated indices, operational costs rise. That makes storage of embedding vectors, and vector-space search, more labor- and infrastructure-intensive. On the other hand, the term-based similarity approach operates on textual metadata already in the index, making it lightweight and fast compared to the other techniques. Overall, while term-based recommenders are lightweight and generally effective, they underperform more advanced methods in click-through rates.

% Results for click-path recommender: 2nd best
% different apporach: models the user next step in the session
% historical session logs vs. recommendation logs
% discuss on second position: second best results....
The click-path recommender (CPR) achieved the second-best performance overall, outperforming the term-based baseline but falling short of embedding-based models. Although CPR also utilizes embeddings (from the nomic model), it differs fundamentally in its objective: rather than recommending items most similar to a target item, it predicts the next item based on users’ historical click paths within a session. In this sense, it models sequential navigation behavior rather than item similarity. Our results suggest that users in the academic domain tend to remain focused on a specific topic and prefer recommendations that are closely related to their current item, rather than following navigation patterns from past sessions. Additionally, CPR was trained on search logs rather than recommendation logs, which may limit its effectiveness in this context. Popularity bias in the training data may further skew its predictions toward frequently clicked items. Prior work has also shown that RNN-based session recommenders struggle in noisy scenarios, such as short sessions \cite{li2024graph} or shifting user intent \cite{wu2019session, liu2018stamp}. Overall, CPR appears less well-suited to our use case.
%Another study demonstrates that supporting sequential models with attention layers improves recommendation effectiveness \cite{thai2022session}. 
%Similar to our findings, Ludewig et al. \cite{ludewig2018evaluation} note that RNN-based recommenders offer an impactful performance gain over the session-based nearest-neighbor approaches.
% long sequences?
% input often consists of a very short sequence of user interactions
%At least for our use case, the click-path recommender was not a convincing recommendation approach for end users.

% larger model
% MOE (ideal for multilingual task)
% finetuned on retrieval task

% Semantic-based recommenders perform best. Why nomic AI? difference in architecture.
Semantic similarity approaches achieved the best performance, consistently attracting more clicks than other methods. Among the three embedding models evaluated, performance varied noticeably, with the nomic model achieving the highest click rates. This may be caused by several factors: (i) it is based on a more recent architecture and the model was introduced lately in 2025, (ii) its Mixture-of-Experts (MoE) design has shown strong performance in multilingual settings \cite{li2025group, ijebu2025ensemble}, and (iii) it has a relatively large size with over 450M parameters. In addition, the model was fine-tuned for retrieval tasks, closely matching our recommendation scenario \cite{nussbaum2025trainingsparsemixtureexperts}.
% mpnet-base vs. minilm: minilm is half the size
In contrast, the two sentence-transformer models (mpnet-base and MiniLM) produced nearly identical results, with no statistically significant difference. Given that MiniLM uses embeddings of half the dimensionality (384 vs. 768), it presents a more efficient model with comparable performance \cite{sentence_transformers}. 
Overall, semantic similarity methods appear simple yet robust in recommending scholarly content.
%Semantic similarity recommenders can be applied to a range of use cases and information types. In our case, we achieved high click-rates for different categories such as publications, variables, research data, and instruments \& tools. One might hypothesize that semantic-similarity recommendations will work across many other domains and information types if users are interested in recommendations based on similarity.

% results across categories
Our category-specific analysis (cp. Table \ref{tab:category-specific}) reveals differences in user interaction across information types. In particular, Publication recommendations receive substantially more clicks than Instruments \& Tools, which is likely associated with differences in category popularity within GESIS Search. As expected, users mainly engage with Publications and Research Data. Interestingly, the term-based recommender performs relatively better in Research Data and Variable categories, even surpassing CPR. This matches findings by Carevic et al. \cite{carevic2020characteristics}, who show that dataset search often resembles known-item search, in which exact-term matching may perform better. While we observe larger fluctuations in performance for the Instruments \& Tools category, the limited number of clicks prevents us from drawing conclusions. Overall, although user behavior varies across categories, the relative ranking of the recommenders remains largely stable.

% missing: discussion on knowledge-based recommenders

% missing: limitation of our approach, e.g., clicks are used as relevance judgments but could also be an indication of (only) simple interest 
%In this study, we adopted a well-recognized evaluation technique, assessing systems by the number of clicks they received from the platform users. Our performance measurement relies on the assumption that user clicks signal relevance between the target and the recommended item. However, they often can indicate only simple interest, accidental clicking, or a quick peek. Therefore, for a more robust evaluation, click counts should be reinforced by other indicators, such as dwell time [REF].
A main limitation of our study is its reliance on click-based evaluation. Our approach assumes that user clicks signal relevance between the target and recommended items; however, clicks may also reflect curiosity, accidental interactions, or brief inspections. To obtain a more robust assessment, click-based metrics should be complemented with additional signals such as dwell time or downstream engagement \cite{sharma2005automated, fox2005evaluating}.

% missing: recommendation for other academic search engines: what are we suggesting to them to deploy good recommendations for end-users?
%Our takeaways from this study are as follows: First, testing a comprehensive set of systems from various classes of recommendation techniques is important to understand which approach suits the use case. Second, the most complex solution may not always be the best for the use case. Sometimes, keeping things simple is enough to achieve decent quality and satisfy users \cite{ludewig2018evaluation}. Finally, taking into account the dynamics of the domain and the type of data being dealt with can be critical to design effective solutions. We suggest considering these points for other academic search systems that plan to deploy recommendations in their platform.
Several takeaways emerge from this work. First, evaluating a diverse set of recommendation approaches is important for identifying the most suitable method for a given use case. Second, more complex models do not necessarily yield better outcomes; simpler methods may provide competitive performance at a fraction of the cost (e.g. semantic similarity vs. CPR)\cite{ludewig2018evaluation}. Finally, domain characteristics and information types can affect users’ information-seeking behavior and should be carefully considered when designing recommender systems. These takeaways may inform the development of recommendation features in other academic search platforms.

% test a system set covering various retrieval or recommendation tecniques
% consider item type
% complex systems are not necessarily better. 

\section{Conclusion}\label{sec_conc}

%missing: one paragraph for the conclusion summarizing the paper very densely.
In this study, we developed and evaluated five different recommender systems for the social science search platform GESIS Search using the STELLA framework. By testing these systems with real users in real-world settings, we identified that semantic similarity models, especially the nomic model, are the most effective at capturing user interest in an academic search scenario.

We also found that the ``best" approach can depend on what is being searched. For Research Data and Variables, simpler term-based methods remain competitive because users in these categories often look for specific, known items. However, for general browsing and publications, users clearly prefer the semantic connections provided by embedding-based models.

In addition to its practical contributions to GESIS Search, this study opens several paths for future work. Planned work includes building a multi-category test collection based on the collected click data, enabling offline evaluation of academic recommender systems in a category-aware setting. This resource will be continuously updated as new interaction data becomes available. Furthermore, we aim to develop new recommendation models customized to our platform, potentially including category-specific approaches. %Finally, we intend to extend our evaluation framework to ad hoc search, extending the scope beyond recommendation.

Overall, this research demonstrates the utility of online evaluation for developing better services and for understanding users’ information-seeking behavior in academic search systems. By exploring diverse techniques and addressing the specific needs of various source types, platforms like GESIS Search can better support researchers in discovering relevant scholarly content.

\begin{credits}
\subsubsection{\ackname} This preprint has not undergone peer review (when applicable) or any post-submission improvements or corrections. The Version of Record of this contribution will be published in TPDL 2026.

\end{credits}

\bibliographystyle{splncs04}
\bibliography{references}

\end{document}